\documentclass[preprint1]{aastex}
\usepackage{amsmath}
\slugcomment{}

\newcommand{\be}{\begin{equation}}
\newcommand{\ee}{\end{equation}}

\makeatletter
\renewcommand{\@make@caption@text}[2]{%
  \begin{center}
    \makebox[\textwidth]{\rmfamily#1.\quad#2}
  \end{center}
}%
\makeatother

\shorttitle{Time Delay Lenses}
\shortauthors{Wei, Wu \& Melia}

\begin{document}

\title{A Comparison of Cosmological Models Using Time Delay Lenses}
\author{Jun-Jie Wei\altaffilmark{1,2}, Xue-Feng Wu\altaffilmark{1,3,4}, and Fulvio Melia\altaffilmark{5}}
\altaffiltext{1}{Purple Mountain Observatory, Chinese Academy of Sciences, Nanjing 210008,
China; jjwei@pmo.ac.cn, xfwu@pmo.ac.cn.}
\altaffiltext{2}{University of Chinese Academy of Sciences, Beijing 100049, China}
\altaffiltext{3}{Chinese Center for Antarctic Astronomy, Nanjing 210008, China.}
\altaffiltext{4}{Joint Center for Particle, Nuclear Physics and Cosmology, Nanjing
University-Purple Mountain Observatory, Nanjing 210008, China.}
\altaffiltext{5}{Department of Physics, The Applied Math Program, and Department of Astronomy,
The University of Arizona, AZ 85721, USA; fmelia@email.arizona.edu.}

\begin{abstract}
The use of time-delay gravitational lenses to examine the cosmological
expansion introduces a new standard ruler with which to test theoretical
models. The sample suitable for this kind of work now includes 12 lens
systems, which have thus far been used solely for optimizing the parameters
of $\Lambda$CDM. In this paper, we broaden the base of support for
this new, important cosmic probe by using these observations to carry out
a one-on-one comparison between {\it competing} models. The currently
available sample indicates a likelihood of $\sim 70-80\%$ that the
$R_{\rm h}=ct$ Universe is the correct cosmology versus $\sim 20-30\%$ for the
standard model. This possibly interesting result reinforces the need to
greatly expand the sample of time-delay lenses, e.g.,
with the successful implementation of the Dark Energy Survey, the VST ATLAS
survey, and the Large Synoptic Survey Telescope. In anticipation of a greatly
expanded catalog of time-delay lenses identified with these surveys, we have
produced synthetic samples to estimate how large they would have to be in
order to rule out either model at a $\sim 99.7\%$ confidence level. We
find that if the real cosmology is $\Lambda$CDM, a sample of $\sim 150$
time-delay lenses would be sufficient to rule out $R_{\rm h}=ct$ at this level of
accuracy, while $\sim 1,000$ time-delay lenses would be required to rule out
$\Lambda$CDM if the real Universe is instead $R_{\rm h}=ct$. This difference
in required sample size reflects the greater number of free parameters available
to fit the data with $\Lambda$CDM.
\end{abstract}


\section{Introduction}
The idea of using gravitational lenses with time delays between the various
images of a background quasar as a cosmological probe was first suggested
by Refsdal (1964). Null geodesics originating with distant variable sources have
different optical paths and pass through dissimilar gravitational potentials.
Their deflection angles and time delays can thus be used to infer lens-system
dependent angular-diameter distances, which can then be compared to theoretical
predictions from general relativity to test the background cosmological expansion
and offer the possibility of testing competing models (see, e.g., Petters et al. 2001;
Schneider et al. 2006).

As of today, time delays have been observed from 21 lensed quasars, a relatively small
subset of the several hundred known strong-lens systems. But this is only the beginning.
In the near future, observational programmes, such as the COSmological MOnitoring of
GRAvItational Lenses (COSMOGRAIL; Eigenbrod et al. 2005) and perhaps also the
International Liquid Mirror Telescope (ILMT) project (Jean et al. 2001), which is now
in the final phases of construction in the Kumaun region of the Himalayas (Sagar et al.
2013), should increase this sample considerably. New strong gravitational lens systems
will also be discovered by cosmic structure surveys, including the Dark Energy
Survey\footnote{One should
take note of the fact, however, that the actual image quality in these
surveys may be inferior to that expected, somewhat mitigating
the possible yield of suitable lens systems for this work. For example, the
DES was aiming to get $0.9^{\prime\prime}$ median FWHM in the r, i, and z band
images for its wide survey. At the end of the first year, the quality is
close to this, though not yet meeting expectations. In addition, g and Y bands
are observed in poorer seeing conditions so their quality is even lower
(Bernstein 2014). This is an important consideration in any discussion
concerning anticipated sample size, given that even SDSS has discovered only
$\sim 5\%$ of the originally expected lens systems.} (DES; Banerji et al.
2008; Buckley-Geer et al. 2014; Schneider 2014), the Large Synoptic Survey
Telescope (LSST; Tyson et al. 2002; Blandford et al. 2006; Marshall et al. 2011;
Chang et al. 2014) project, and the VST ATLAS survey (Koposov et al.
2014), and time delays will be measured for a sub-sample of these with subsequent
monitoring observations. Oguri \& Marshall (2010) carried out a detailed calculation of
the likely yields of several planned surveys, using realistic distributions for the lens and
source properties and taking magnification bias and image configuration detectability into
account. They found that upcoming wide-field synoptic surveys should detect several thousand
lensed quasars. In particular, LSST should find more than $\sim$8,000 lensed quasars,
some 3,000 of which will have well-measured time delays.

Several attempts have already been made to demonstrate the usefulness of these
data for constraining the cosmological parameters in the standard model,
$\Lambda$CDM (see, e.g., Paraficz \& Hjorth 2009, 2010;  Balm\`es \& Corasaniti
2013; and Suyu et al. 2013). In this paper, we broaden the base
of support for this cosmic probe by demonstrating its usefulness in testing
{\it competing} cosmological models. In particular, we show that the currently
available sample of time-delay lensing systems favors the $R_{\rm h}=ct$ Universe
with a likelihood of $\sim 70-80\%$ of being correct, versus $\sim 20-30\%$
for $\Lambda$CDM. Though this result is still only marginal, it nonetheless
calls for a significant increase in the sample of suitable lensing systems in
order to carry out more sophisticated and higher precision measurements.

In \S~2, we describe the key theoretical steps used in the application of time-delay
lenses for cosmological testing, and we apply this procedure to the currently known
sample of 12 systems in \S~3. We discuss the results of our one-on-one comparison
between $\Lambda$CDM and $R_{\rm h}=ct$ in \S~4. As we shall see, model selection
tools favor the latter, but the distinction, given the relatively small number of
lenses, is still not strong enough to completely rule out either model.
We therefore estimate the sample size
required from future surveys to reach likelihoods of $\sim 99.7\%$ and $\sim0.3\%$,
for a $3\sigma$ confidence limit, and we present our conclusions in \S~6.

\section{Strong Lensing}
The measurement of time delays in strong gravitational lenses is not straightforward,
due in part to the uncertainty associated with the lens mass distribution and the
possible presence of other perturbing masses along the line-of-sight. To this point,
two principal methods have been employed to model the lens itself, which may be
characterized as either simple-parametric (see, e.g., Oguri et al. 2002; Keeton et al.
2003) or grid-based parametric (see, e.g., Warren \& Dye 2003; Bradac et al. 2008;
Coles 2008; Suyu et al. 2013) approaches. The former uses simply-parametrized forms
for the mass distribution of the deflector, while the latter uses as parameters a
grid of pixels, to describe either the potential or the mass distribution of the
deflector, and/or the source surface brightness distribution. Some have also used
a hybrid approach, in which pixellated corrections were made to a simply
parametrized mass model (Suyu et al. 2010; Vegetti et al. 2010).

The grid-based methods are regularized, often by imposing physical priors, otherwise
they would just fit the noise. The simply parametrized methods can even be computationally
more intensive, depending on the choice of the parameters. If an appropriate
sub-sample of homogeneous systems can be identified---meaning a set of lenses
whose properties provide evidence that the same lens model description may be
used with the same level of statistical significance---the simply parametrized
method can serve as an ideal first attempt at gauging whether the
image-inversion effort is warranted with follow-up analysis. This
is the method we will be using in this paper.

For a given image $i$ at angular position $\vec{\theta}_i$, with the source position
at angle $\vec{\beta}$, the time delay, $\Delta t_i$, is the combined effect of the
difference in path length between the straight and deflected rays, and the gravitational
time dilation for the ray passing through the effective gravitational potential of
the lens, $\Psi(\vec{\theta}_i)$:
\begin{equation}
\Delta t_i = {1+z_l\over c}{D_A(0,z_s)D_A(0,z_l)\over D_A(z_l,z_s)}\left[{1\over 2}
(\vec{\theta}_i-\vec{\beta})^2-\Psi(\vec{\theta}_i)\right]
\end{equation}
(see, e.g., Blandford \& Narayan 1986, and references cited therein), where
$z_l$ and $z_s$ are the lens and source redshifts, respectively, and
$D_A(z_1,z_2)$ is the angular-diameter distance between redshifts $z_1$ and
$z_2$. If the lens geometry $\vec{\theta}_i-\vec{\beta}$ and the lens potential
$\Psi$ are known, the time delay measures the ratio
\begin{equation}
\mathcal{R}\equiv {D_A(0,z_s)D_A(0,z_l)\over D_A(z_l,z_s)}\;,
\end{equation}
also known as the time-delay distance, which depends on the
cosmological model.

It has been known for over a decade that lens spiral and elliptical galaxies
have a mass distribution that is well approximated by power-law density profiles
(Witt et al. 2000; Rusin et al. 2003), for which the lens potential may be written
\begin{equation}
\Psi(\vec{\theta})={b^2\over 3-n}\left({\theta\over b}\right)^{3-n}\;,
\end{equation}
in terms of the deflection scale $b$ and index $n$. The single isothermal sphere
(SIS) is the special
case corresponding to $n=2$, for which $b=4\pi D_A(z_l,z_s)\sigma^2_\star/D_A(0,z_s)$,
where $\sigma_\star$ is the velocity dispersion of the lensing galaxy. Though some
systems have shallow profiles with $n<1$, measurements of galaxy density
distributions suggest that $n$ is generally close to the isothermal value.
Thus, in addition to the SIS model being convenient for its simplicity,
it is actually also a surprisingly useful and accurate model for lens
galaxies (Guimaraes \& Sodr\'e 2009; Koopmans et al. 2009). And for such systems with
only two images at $\vec{\theta_A}$ and $\vec{\theta_B}$, the time delay
is given by the expression
\begin{equation}
\Delta t=t_A-t_B={1+z_l\over 2c}\mathcal{R}(z_l,z_s)\left(\theta_B^2-\theta_A^2\right)\;.
\end{equation}

Treu et al. (2006) found that the ratio $f\equiv\sigma_{\star}/\sigma_{\rm SIS}$
is very close to unity, where $\sigma_{\rm SIS}$ includes systematic errors
in the rms deviation of the velocity dispersion and a softened isothermal sphere potential
(see additional details below). Note that if the velocity dispersion $\sigma_\star$ of the lensing galaxy could
also be observed, and assuming a ratio $f=1$, two of the angular-diameter distances appearing in equation~(2)
could be replaced with the measured value of $\sigma_\star$ and the Einstein radius
$\theta_{\rm E}=(\theta_A+\theta_B)/2$. An alternative approach would be to impose
some prior (see, e.g., Oguri 2007), or to compute $\sigma_\star$ from a
dynamical model (see, e.g., Treu \& Koopmans 2002). As shown by Paraficz \& Hjorth (2009),
the quantity $\Delta t/\sigma^2_{\rm SIS}$ is more sensitive to the cosmological
parameters than $\Delta t$ or $\sigma^2_{\rm SIS}$ separately, so this additional
datum would improve the reliability with which this approach could distinguish
between competing models. As of today, however, there are simply too few time-delay
lenses with the corresponding measure of $\sigma_\star$, so all of the analysis
we carry out in this paper will be based solely on the use of equation~(4).
Even looking to the future, velocity dispersions are particularly difficult
to measure for these systems precisely because they are crowded by quasars
that make the time delay measurement possible.

In $\Lambda$CDM, the angular-diameter distance depends
on several parameters, including $H_0$ and the mass fractions
$\Omega_{\rm m} \equiv \rho_{\rm m}/\rho_{\rm c}$, $\Omega_{\rm r}\equiv
\rho_{\rm r}/\rho_{\rm c}$, and $\Omega_{\rm de}\equiv \rho_{\rm de}/
\rho_{\rm c}$, defined in terms of the current matter ($\rho_{\rm m}$),
radiation ($\rho_{\rm r}$), and dark energy ($\rho_{\rm de}$) densities,
and the critical density $\rho_{\rm c}\equiv 3c^2H_0^2/8\pi G$.
Assuming zero spatial curvature, so that $\Omega_{\rm m}+\Omega_{\rm r}
+\Omega_{\rm de}=1$, the angular-diameter distance between redshifts
$z_1$ and $z_2$ ($>z_1$) is given by the expression
\begin{equation}
D_A^{\Lambda{\rm CDM}}(z_1,z_2)={c\over H_0}{1\over (1+z_2)}\int_{z_1}^{z_2}
\left[\Omega_{\rm m}(1+z)^3+\Omega_{\rm r}(1+z)^4+\Omega_{\rm de}
(1+z)^{3(1+w_{\rm de})}\right]^{-1/2}\;dz\;,
\end{equation}
where $p_{\rm de}=w_{\rm de}\rho_{\rm de}$ is the dark-energy equation
of state. Thus, the essential free parameters in flat $\Lambda$CDM are
$H_0$, $\Omega_{\rm m}$ and $w_{\rm de}$ (since radiation is insignificant
at gravitational lensing redshifts).
In the $R_{\rm h}=ct$ Universe (Melia 2007; Melia \& Shevchuk 2012),
the angular-diameter distance depends only on $H_0$. In this cosmology,
\begin{equation}
D_A^{R_{\rm h}=ct}(z_1,z_2)={c\over H_0}{1\over (1+z_2)}
\ln\left({1+z_2\over 1+z_1}\right)\;.
\end{equation}

\section{Sample of Time-Delay (Two-image) Lensing Systems}

In their careful Bayesian approach to constraining $H_0$ within the framework
of $\Lambda$CDM, Balm\`es \& Corasaniti (2013) collected a sample of time-delay
lenses for which Bayesian selection techniques can identify the lens mass model
with the highest probability of describing the lens system. Rather than attempting
to model individual lenses in all their complexity, the goal was to identify the
model whose parameters significantly influence the time-delay, allowing them to
average over individual mass model parameter uncertainties on a homogeneous mass
sample. The first selection criterion in such an approach is therefore a restriction
to two-image lenses, listed in Table 1, which seem to be more likely than other lens
systems to be consistent with a simple power-law (or even SIS) profile. Paraficz
\& Hjorth (2010) followed the alternative method of using inversion techniques
with each individual intensity image to map the mass distribution in each
individual lens system, and produced a very useful comparison of their mass
profiles, shown in Figure~1 of that paper. Indeed, the two-image lenses are
significantly more symmetric than the rest. But though the object constituting
the lens has been identified unambiguously in all the entries listed in Table 1,
it is not yet clear whether these are part of a group or whether perturbators
contribute along the line-of-sight. Thus, at this stage, an important caveat to our
conclusions is that external shear may yet be contributing to some selection bias.

\begin{deluxetable}{lllccccccl}
\tablewidth{476pt}
\tabletypesize{\footnotesize}
\tabletypesize{\tiny}
\tablecaption{Time Delay (Two-image) Lenses}\tablenum{1}
\tablehead{{\rm System}&\colhead{$z_l$}&\colhead{$z_s$}&\colhead{$\theta_A$}&\colhead{$\theta_B$}&
\colhead{$\Delta t=t_A-t_B$}&\colhead{$\mathcal{R}_{\rm obs}$}&\colhead{$\mathcal{R}_{\Lambda{\rm CDM}}$}&
\colhead{$\mathcal{R}_{R_{\rm h}=ct}$}& {\rm Refs.} \\
&&&({\rm arcsec})&({\rm arcsec})&(days)&(Gpc)&(Gpc)&(Gpc)&
} \startdata
{\rm	B}0218+357	&	0.685	&	0.944	&	$0.057\pm0.004$	&	$0.280\pm0.008$	&	$+10.5\pm0.2$	&	$5.922	\pm	1.757$	&	5.268	&	5.361	& 1--3 \\
{\rm	B}1600+434	&	0.414	&	1.589	&	$1.14\pm0.075$	&	$0.25\pm0.074$	&	$-51.0\pm2.0$	&	$2.082	\pm	0.677$	&	1.403	&	1.435	& 4,5 \\
{\rm	FBQ}0951+2635	&	0.26	&	1.246	&	$0.886\pm0.004$	&	$0.228\pm0.008$	&	$-16.0\pm2.0$	&	$1.237	\pm	0.391$	&	0.917	&	0.956	& 6 \\
{\rm	HE}1104-1805	&	0.729	&	2.319	&	$1.099\pm0.004$	&	$2.095\pm0.008$	&	$152.2\pm3.0$	&	$1.976	\pm	0.575$	&	2.202	&	2.170	& 2,7,8\;\; \\
{\rm	HE}2149-2745	&	0.603	&	2.033	&	$1.354\pm0.008$	&	$0.344\pm0.012$	&	$-103.0\pm12.0$	&	$2.676	\pm	0.837$	&	1.909	&	1.908	& 6,9 \\
{\rm	PKS}1830-211	&	0.89	&	2.507	&	$0.67\pm0.08$	&	$0.32\pm0.08$	&	$-26\pm5$	&	$2.835	\pm	1.385$	&	2.620	&	2.546	& 10,11 \\
{\rm	Q}0142-100	&	0.49	&	2.719	&	$1.855\pm0.002$	&	$0.383\pm0.005$	&	$-89\pm11$	&	$1.295	\pm	0.408$	&	1.428	&	1.431	& 6,12 \\
{\rm	Q}0957+561	&	0.36	&	1.413	&	$5.220\pm0.006$	&	$1.036\pm0.11$	&	$-417.09\pm0.07$	&	$0.837	\pm	0.243$	&	1.256	&	1.294	& 6, 13,14\;\; \\
{\rm	SBS} 0909+532	&	0.83	&	1.377	&	$0.415\pm0.126$	&	$0.756\pm0.152$	&	$+45.0\pm5.5$	&	$4.398	\pm	3.107$	&	4.085	&	4.072	& 6,15 \\
{\rm	SBS} 1520+530	&	0.717	&	1.855	&	$1.207\pm0.004$	&	$0.386\pm0.008$	&	$-130.0\pm3.0$	&	$4.135	\pm	1.203$	&	2.435	&	2.419	& 6,16 \\
{\rm	SDSS J}1206+4332	&	0.748	&	1.789	&	$1.870\pm0.088$	&	$1.278\pm0.097$	&	$-116\pm5$	&	$2.543	\pm	0.934$	&	2.632	&	2.612	& 17 \\
{\rm	SDSS J}1650+4251	&	0.577	&	1.547	&	$0.872\pm0.027$	&	$0.357\pm0.042$	&	$-49.5\pm1.9$	&	$3.542	\pm	1.082$	&	2.079	&	2.098	& 6,18 \\
\enddata
\tablenotetext{References:\hskip0.2in} {(1) Carilli et al. (1993); (2) Leh\'ar et al. (2000);
(3) Wucknitz et al. (2004); (4) Jackson et al. (1995); (5) Dai \& Kochanek (2005);
(6) Kochanek et al. (2008); (7) Wisotzki et al. (1993); (8) Poindexter et al. (2007);
(9) Burud et al. (2002); (10) Lovell et al. (1998); (11) Meylan et al. (2005);
(12) Koptelova et al. (2012); (13) Falco et al. (1997); (14) Colley et al. (2003);
(15) Dai \& Kochanek (2009); (16) Auger et al. (2008); (17) Paraficz et al. (2009);
(18) Vuissoz et al. (2007).}
\end{deluxetable}

Note, however, that the two-image criterion may not be sufficient to
guarantee a simple power-law lens model. Balm\`es and Corasaniti (2013)
concluded from this sample that nine have Bayes factors favoring such a
mass profile, though six of these are somewhat inconclusive. Thus, for a
more balanced assessment in our analysis, we will consider two sample cuts,
one with the full set of 12 two-image lenses listed in Table 1, and
the second with just these nine: B1600+434, SBS 1520+530, SDSS J1650+4251,
B0218+357, FBQ 0951+2635, HE 2149-2745, PKS 1830-211, Q0142-100,
and SBS 0909+532.

For each model, we find the optimized fit by maximizing the joint likelihood function
\begin{equation}
\begin{split}
L(\sigma_{\rm SIS},\xi)\propto\prod_{i=1}
\frac{1}{\sqrt{\sigma^{2}_{\rm SIS}+\sigma^{2}_{\mathcal{R}_{i}}}}\times
\exp\left[-\frac{\left(\mathcal{R}_{\rm th, \emph{i}}[\xi]-\mathcal{R}_{\rm obs,
\emph{i}}\right)^{2}} {2(\sigma^{2}_{\rm SIS}+\sigma^{2}_{\mathcal{R}_{i}})} \right]\;,
\end{split}
\end{equation}
where `th' stands for either $\Lambda$CDM or $R_{\rm h}=ct$, $\mathcal{R}_{\rm th}$ is
the theoretical time-delay distance calculated from $z_{l,i}$, $z_{s,i}$
and the model specific parameters $\xi$,
$\mathcal{R}_{\rm obs}$ is the measured value, and $\sigma_{\mathcal{R}}$ is the
dispersion of $\mathcal{R}_{\rm obs}$. The measured time-delay distance is
\begin{equation}
\mathcal{R}_{\rm obs}(z_l,z_s)= {2c \over 1+z_l}{\Delta t \over (\theta_B^2-\theta_A^2)},
\end{equation}
so the propagated error $\sigma_{\mathcal{R}}$ in $\mathcal{R}_{\rm obs}$ is
\begin{equation}
\sigma_{\mathcal{R}}=\mathcal{R}_{\rm obs}\left[\left({\sigma_{\Delta t}\over \Delta t}\right)^{2}
+4\left({\theta_B\sigma_{\theta_B}\over \theta_B^2-\theta_A^2}\right)^2
+4\left({\theta_A\sigma_{\theta_A}\over \theta_B^2-\theta_A^2}\right)^2\right]^{1/2}\;.
\end{equation}
The dispersion $\sigma_z$ in the measured redshifts $z_l$ and $z_s$ (which enter
through the angular distances in $\mathcal{R}_{\rm obs}$) will be ignored here
because a careful analysis of SDSS quasar spectra shows that $\sigma_z/(1+z)\sim 10^{-4}$
(Hewett \& Wild 2010), much smaller than the other errors appearing in equation~(9).

However, we must include another source of error, in addition to $\sigma_{\mathcal{R}}$,
that we will call $\sigma_{\rm SIS}$; this takes into account at least
several effects that apparently give rise to the observed scatter of individual
lenses about the pure SIS profile. These include: systematic errors in the rms
deviation of the velocity dispersion and a softened isothermal sphere potential,
which tends to decrease the typical image separations. Moreover, Koopmans et
al. (2009) showed that the mean mass density profile is slightly steeper than SIS
and has significant scatter, not to mention that the line of sight contribution
is generally non-zero on average (Suyu et al. 2010).
According to Cao et al. (2012), $\sigma_{\rm SIS}$ may be as big as
$\sim 20\%$, depending on how many such factors actually contribute to this
scatter. In our approach, we will adopt the additional free parameter $\eta$
to relate the dispersion $\sigma_{\rm SIS}$ to the measured effective
lensing distance $\mathcal{R}_{\rm obs}$, according to
\begin{equation}
\sigma_{\rm SIS}\equiv  \eta\mathcal{R}_{\rm obs}.
\end{equation}
We will add $\sigma_{\rm SIS}$ and $\sigma_{\mathcal{R}}$ in quadrature,
and optimize the parameter $\eta$ for each fit individually though, as we shall
see, the value of $\eta$ appears to be quite independent of the model itself. Thus,
the total uncertainty $\sigma_{\rm tot}$ in $\mathcal{R}_{\rm obs}$ is given by
the expression $\sigma_{\rm tot}^{2}=\sigma_{\rm SIS}^{2}+\sigma_{\mathcal{R}}^{2}$.

\section{Results and Discussion}
We have used the data shown in Table~1 to directly compare $\Lambda$CDM with
the $R_{\rm h}=ct$ Universe. The parameters in each model were individually
optimized by maximizing the likelihood estimation, as described above. We will
use two well established priors associated with the concordance $\Lambda$CDM
model: (i) dark energy is a cosmological constant, so $w_{\rm de}=-1$; and the
spatial curvature constant is zero, i.e., the Universe is flat, so that
$\Omega_{\Lambda}=1-\Omega_{\rm m}$. But to allow for added flexibility in
the optimization of the model fit, we will employ three free parameters:
$H_{0}$, $\Omega_{\rm m}$, and $\eta$. We have restricted the fraction
$\Omega_{\rm m}$ to the range $(0.0,1.0)$. With the full sample
of 12 time-delay lenses, $\Lambda$CDM fits the data with a maximum likelihood
for $H_{0}=87^{+17}_{-16}$ ($1\sigma$) km $\rm s^{-1}$ $\rm Mpc^{-1}$,
$\Omega_{\rm m}=0.48^{+0.25}_{-0.37}$ ($1\sigma$) and $\eta=0.29^{+0.15}_{-0.09}$ ($1\sigma$). The best fit with the
$R_{\rm h}=ct$ Universe has only two free parameters: $H_{0}=81_{-13}^{+18}$ ($1\sigma$)
km $\rm s^{-1}$ $\rm Mpc^{-1}$ and $\eta=0.29^{+0.16}_{-0.09}$ ($1\sigma$).
The entries in column~7 of Table~1 are calculated from the observed angles and
time delays. By comparison, columns~8 and 9 show the entries for
$\mathcal{R}_{\Lambda{\rm CDM}}$ and $\mathcal{R}_{R_{\rm h}=ct}$,
respectively, corresponding to these best-fit parameters using all
12 lens systems.

\begin{figure}[ht]
\centerline{\includegraphics[angle=0,scale=0.8]{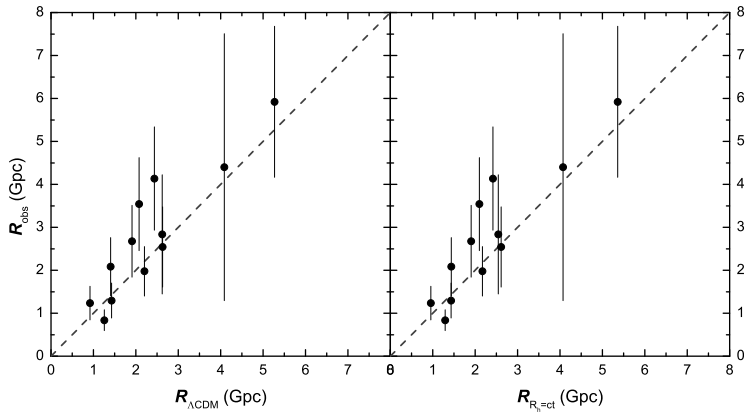}}
\vskip-0.2in
\caption{Twelve $\mathcal{R}$ measurements, with error bars, compared to two
theoretical models: (\emph{left}) the standard $\Lambda$CDM cosmology,
assuming a flat universe, with $H_{0}=87^{+17}_{-16}$ km $\rm s^{-1}$ $\rm Mpc^{-1}$,
$\Omega_{\rm m}=0.48^{+0.25}_{-0.37}$ and $\eta=0.29^{+0.15}_{-0.09}$; and (\emph{right}) the $R_{\rm h}=ct$
Universe, with $H_{0}=81_{-13}^{+18}$ km $\rm s^{-1}$ $\rm Mpc^{-1}$
and $\eta=0.29^{+0.16}_{-0.09}$.}
\end{figure}

To facilitate a direct comparison between $\Lambda$CDM and $R_{\rm h}=ct$,
we show in Figure~1 the 12 observed values of $\mathcal{R}_{\rm obs}$
compared with those predicted by these two theoretical models.
The optimized values of $\eta$ and the maximum likelihood
are quite similar for these two cases.
However, these models formulate their observables (such as the angular
diameter distances in Equations~5 and 6) differently, and do not have
the same number of free parameters. Therefore a comparison of the
likelihoods for either being closer to the `true' model must be based
on model selection tools.

\begin{figure}[ht]
\centerline{\includegraphics[angle=0,scale=0.8]{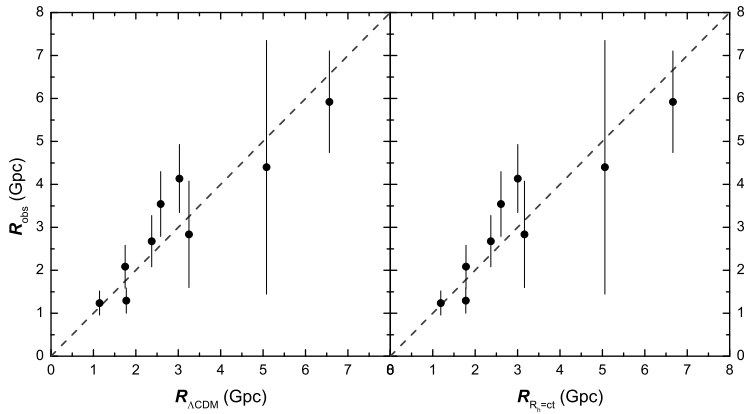}}
\vskip-0.2in
\caption{Same as Figure~1, except now for the reduced sample of 9
lens systems. The optimized $\Lambda$CDM model has $H_{0}=69^{+12}_{-11}$ km
$\rm s^{-1}$ $\rm Mpc^{-1}$, $\Omega_{\rm m}=0.51^{+0.32}_{-0.27}$ and $\eta=0.19^{+0.16}_{-0.07}$. The optimized
$R_{\rm h}=ct$ Universe has $H_{0}=65_{-8.8}^{+13}$ km $\rm s^{-1}$
$\rm Mpc^{-1}$ and $\eta=0.19^{+0.16}_{-0.08}$.}
\end{figure}

Several information criteria commonly used in cosmology (see,
e.g., Melia \& Maier 2013, and references cited therein) include
the Akaike Information Criterion, ${\rm AIC}\equiv-2\ln L+2n$, where
$L$ is the maximum likelihood,
$n$ is the number of free parameters (Liddle 2007),
the Kullback Information Criterion, ${\rm KIC}=-2\ln L+3n$ (Cavanaugh
et al. 2004), and the Bayes Information Criterion,
${\rm BIC}=-2\ln L+(\ln N)n$, where $N$ is the number of data points
(Schwarz et al. 1978). With ${\rm AIC}_\alpha$ characterizing model $\mathcal{M}_\alpha$,
the unnormalized confidence that this model is true is the Akaike
weight $\exp(-{\rm AIC}_\alpha/2)$. Model $\mathcal{M}_\alpha$ has likelihood
\begin{equation}
P(\mathcal{M}_\alpha)= \frac{\exp(-{\rm AIC}_\alpha/2)}
{\exp(-{\rm AIC}_1/2)+\exp(-{\rm AIC}_2/2)}
\end{equation}
of being the correct choice in this one-on-one comparison. Thus, the difference
$\Delta \rm AIC \equiv {\rm AIC}_2\nobreak-{\rm AIC}_1$ determines the extent to which $\mathcal{M}_1$
is favoured over~$\mathcal{M}_2$. For Kullback
and Bayes, the likelihoods are defined analogously.
For the two optimized fits discussed above, the magnitude of the difference
$\Delta \rm AIC={\rm AIC}_2\nobreak-{\rm AIC}_1=1.7$, indicates that
$\mathcal{M}_1$ is to be preferred over $\mathcal{M}_2$. According to Equation~(11),
the likelihood of $R_{\rm h}=ct$ (i.e. $\mathcal{M}_1$) being the correct choice
is $P(\mathcal{M}_1)\approx 70\%$. For $\Lambda$CDM (i.e. $\mathcal{M}_2$),
the corresponding value is $P(\mathcal{M}_2)\approx 30\%$. With the alternatives
KIC and BIC, the magnitude of the differences $\Delta \rm KIC={\rm KIC}_2\nobreak-{\rm KIC}_1=2.7$
and $\Delta \rm BIC={\rm BIC}_2\nobreak-{\rm BIC}_1=2.2$, indicates that $R_{\rm h}=ct$
is favored over $\Lambda$CDM by a likelihood of $\approx 75\%-80\%$ versus $20\%-25\%$.

We also carried out a one-on-one comparision using the reduced sample
of only 9 two-image lens systems. In this case, the best $\Lambda$CDM
fit has a maximum likelihood for $H_{0}=69^{+12}_{-11}$ ($1\sigma$)
km $\rm s^{-1}$ $\rm Mpc^{-1}$, $\Omega_{\rm m}=0.51^{+0.32}_{-0.27}$ ($1\sigma$)
and $\eta=0.19^{+0.16}_{-0.07}$ ($1\sigma$). For $R_{\rm h}=ct$, the best fit corresponds to
$H_{0}=65_{-8.8}^{+13}$ ($1\sigma$) km $\rm s^{-1}$ $\rm Mpc^{-1}$
and $\eta=0.19^{+0.16}_{-0.08}$ ($1\sigma$). Figure~2
is similar to Figure~1, except now for the reduced sample of 9 lenses.
In this case, the magnitude of the differences $\Delta \rm AIC=2.0$, $\Delta \rm KIC=3.0$,
and $\Delta \rm BIC=2.2$, indicates that $R_{\rm h}=ct$ is preferred over $\Lambda$CDM with a
likelihood of $\approx 73\%$ versus $\approx 27\%$ using AIC,
$\approx 82\%$ versus $\approx 18\%$ using KIC, and $\approx 75\%$
versus $\approx 25\%$ using BIC.

\section{Monte Carlo Simulations with a Synthetic Sample}
Our results in this paper have shown that time-delay lenses can in fact
be used to select one model over another in a one-on-one comparison.
But though the likelihood of $R_{\rm h}=ct$ being closer to the
correct cosmology than $\Lambda$CDM is $\sim 80\%$ or higher, depending
on the choice of information criterion, the outcome $\Delta\equiv$
AIC$_1-$ AIC$_2$ (and analogously for KIC and BIC) is judged `positive'
in the range $\Delta=2-6$, and `strong' for $\Delta>6$. The constraints
based on the currently known 12 lens systems should therefore be
characterized as `positive,' though not yet strong. In this section, we
will estimate the sample size required to significantly strengthen the
evidence in favor of $R_{\rm h}=ct$ or $\Lambda$CDM, by conservatively
seeking an outcome even beyond $\Delta=6$, i.e., we will see what is
required to produce a likelihood $\sim 99.7\%$ versus $\sim 0.3\%$,
corresponding to $3\sigma$.

We will consider two cases: one in which the background cosmology is
assumed to be $\Lambda$CDM, and a second in which it is $R_{\rm h}=ct$,
and we will attempt to estimate the number of time-delay lenses
required in each case in order to rule out the alternative (incorrect)
model at a $\sim 99.7\%$ confidence level. The synthetic time-delay
lenses are each characterized by a set of parameters denoted as
($z_{l}$, $z_{s}$, $\Delta t$, $\Theta$), where $\Theta\equiv\theta_B^2-
\theta_A^2$, and are generated using the following procedure:

1. Since the 12 observed lens redshifts all fall in the range $0.26\leq z_{l} \leq 0.89$,
and the source redshifts are $1.246\leq z_{s} \leq 2.719$, with a time delay
$-150\le \Delta t \le 150$ (days), we assign $z_{l}$ uniformly between $0.2$
and $1.0$, $z_{s}$ uniformly between $1.2$ and $3.0$, and $\Delta t$ uniformly
between $-150$ and $150$ days.

2. We first infer $\Theta\equiv(\theta_B^2-\theta_A^2)$
from Equation~4 corresponding either to the $R_{\rm h}=ct$ Universe with
$H_{0}=70$ km $\rm s^{-1}$ $\rm Mpc^{-1}$ (\S~5.1), or to
$\Lambda$CDM with $\Omega_{\rm m}=0.3$, $\Omega_{\Lambda}=0.7$ and
$H_{0}=70$ km $\rm s^{-1}$ $\rm Mpc^{-1}$ (\S~5.2).
We then assign a random deviation ($\Delta \Theta$)
to the $\Theta$ value within the $3\sigma$ confidence interval, i.e., we put
$\Theta'=\Theta+(2x-1)\cdot3\sigma$, where $x$ is a uniform random variable
between $0$ and $1$, and $\sigma=0.04\;\Theta$. This value of $\sigma$ is
taken from the current observed sample, which shows a median deviation
$\sim 0.04\,\Theta$. The same relative error is assigned to $\Theta'$.

3. Since the observed $\sigma_{\Delta t}$ is about $4\%$ of $\Delta t$,
we will also assign dispersion $\sigma_{\Delta t}=0.04\;\Delta t$ to the
synthetic sample.

This sequence of steps is repeated for each lens system in the sample, which
is enlarged until the likelihood criterion discussed above is reached. As with
the real 12-lens sample, we optimize the model fits by maximizing the joint
likelihood function in equation~(7). We employ Markov-chain Monte Carlo techniques.
In each Markov chain, we generate $10^5$ samples according to the likelihood function.
Then we derive the coefficients $\eta$ and the cosmological parameters from a
statistical analysis of the sample.

\begin{figure}[hp]
\centerline{\includegraphics[angle=0,scale=1.0]{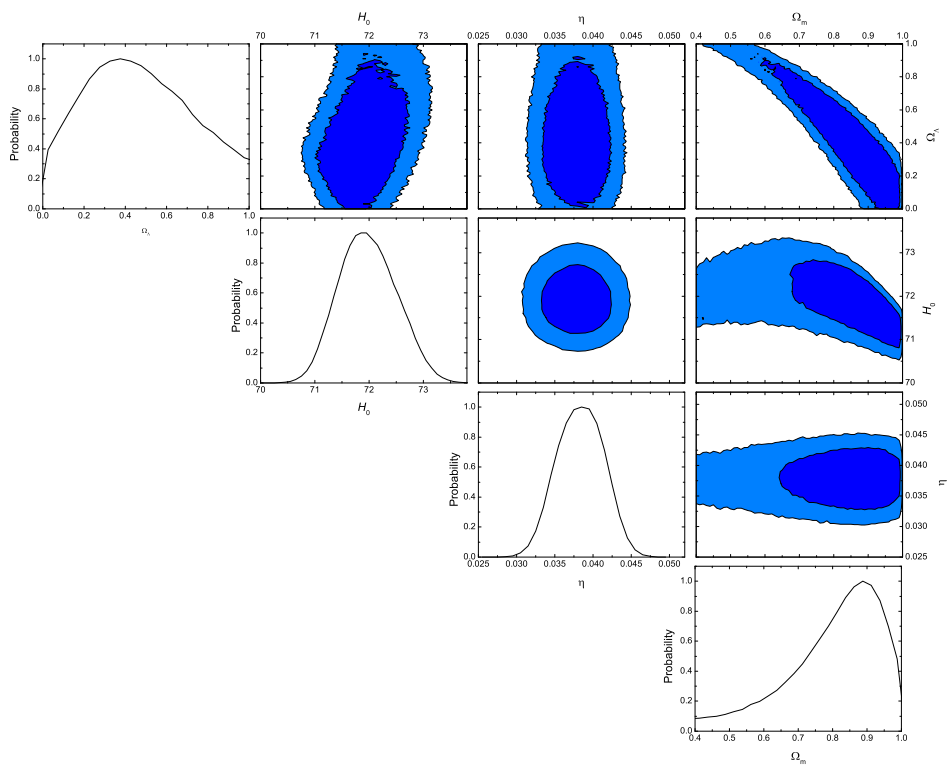}}
\vskip-0.2in
\caption{The 1-D probability distributions and 2-D regions with the $1\sigma$ and
$2\sigma$ contours corresponding to the parameters $\Omega_{\rm m}$, $\Omega_{\Lambda}$,
$H_{0}$, and $\eta$ in the best-fit $\Lambda$CDM model, using the simulated sample with
1,000 lens systems, assuming $R_{\rm h}=ct$ as the background cosmology.}
\end{figure}

\subsection{Assuming $R_{\rm h}=ct$ as the Background Cosmology}
We have found that a sample of at least 1,000 time-delay lenses is required
in order to rule out $\Lambda$CDM at the $\sim 99.7 \%$ confidence level. The
optimized parameters corresponding to the best-fit $\Lambda$CDM model for
these simulated data are displayed in figure~3. To allow for the greatest
flexibility in this fit, we relax the assumption of flatness, and allow
$\Omega_\Lambda$ to be a free parameter, along with $\Omega_{\rm m}$. Figure~3
shows the 1-D probability distribution for each parameter ($\Omega_{\rm m}$,
$\Omega_{\Lambda}$, $H_{0}$, $\eta$), and 2-D plots of the $1\sigma$ and
$2\sigma$ confidence regions for two-parameter combinations. The best-fit
values for $\Lambda$CDM using the simulated sample with 1,000 lens systems
in the $R_{\rm h}=ct$ Universe are $\Omega_{\rm m}=0.85_{-0.21}^{+0.21}$ $(1\sigma)$,
$\Omega_{\Lambda}=0.48_{-0.48}^{+0.47}$, $H_{0}=72_{-0.80}^{+0.81}$ $(1\sigma)$ km
$\rm s^{-1}$ $\rm Mpc^{-1}$, and $\eta=0.038_{-0.0050}^{+0.0040}$ $(1\sigma)$.

\begin{figure}[hp]
\centerline{\hskip 0.5in\includegraphics[angle=0,scale=0.8]{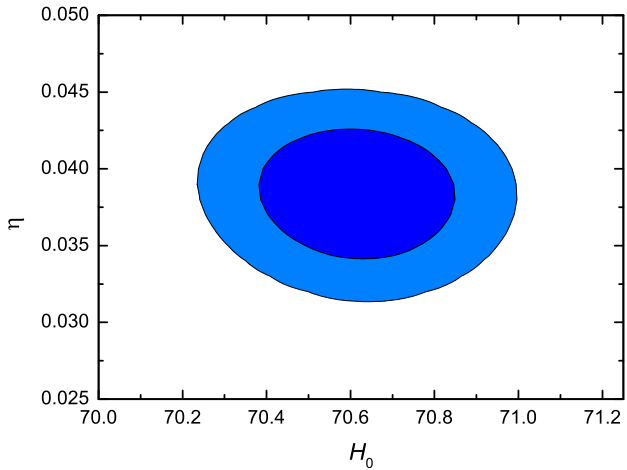}}
\vskip-0.2in
\caption{The 2-D region with the $1\sigma$ and $2\sigma$
contours for the parameters $H_{0}$ and $\eta$ in the $R_{\rm h}=ct$ Universe,
using a sample of 1,000 time-delay lenses, simulated with $R_{\rm h}=ct$ as the
background cosmology. The assumed value for $H_0$ in the simulation was
$H_{0}=70$ km $\rm s^{-1}$ $\rm Mpc^{-1}$.}
\end{figure}

In figure~4, we show the corresponding 2-D contours in the $H_{0}-\eta$ plane for
the $R_{\rm h}=ct$ Universe. The best-fit values for the simulated sample are
$H_{0}=71_{-0.24}^{+0.24}$ $(1\sigma)$ km $\rm s^{-1}$ $\rm Mpc^{-1}$ and
$\eta=0.038_{-0.0040}^{+0.0050}(1\sigma)$.

Since the number $N$ of data points in the sample is now much greater than one, the
most appropriate information criterion to use is the BIC. The logarithmic penalty
in this model selection tool strongly suppresses overfitting if $N$ is large
(the situation we have here, which is deep in the asymptotic regime). With $N=1,000$,
our analysis of the simulated sample shows that the BIC would favor the $R_{\rm h}=ct$
Universe over $\Lambda$CDM by an overwhelming likelihood of $99.7\%$ versus only $0.3\%$
(i.e., the prescribed $3\sigma$ confidence limit).

\subsection{Assuming $\Lambda$CDM as the Background Cosmology}
In this case, we assume that the background cosmology is $\Lambda$CDM,
and seek the minimum sample size to rule out $R_{\rm h}=ct$ at the
$3\sigma$ confidence level. We have found that a minimum of 135 time-delay
lenses are required to achieve this goal. To allow for the greatest flexibility
in the $\Lambda$CDM fit, here too we relax the assumption of flatness, and
allow $\Omega_\Lambda$ to be a free parameter, along with $\Omega_{\rm m}$.
In figure~5, we show the 1-D probability distribution for each parameter
($\Omega_{\rm m}$, $\Omega_{\Lambda}$, $H_{0}$, $\eta$), and 2-D plots of
the $1\sigma$ and $2\sigma$ confidence regions for two-parameter
combinations. The best-fit values for $\Lambda$CDM using this simulated sample
with 135 lens systems are $\Omega_{\rm m}=0.34_{-0.18}^{+0.20}$ $(1\sigma)$,
$\Omega_{\Lambda}=0.58$, $H_{0}=71_{-2.1}^{+2.1}$ $(1\sigma)$ km
$\rm s^{-1}$ $\rm Mpc^{-1}$, and $\eta=0.041_{-0.013}^{+0.011}$ $(1\sigma)$.
Note that the simulated lenses give a good constraint on $\Omega_{\rm m}$, but
a weak one on $\Omega_{\Lambda}$; only an upper limit of 0.90 can be set at
the $1\sigma$ confidence level.

The corresponding 2-D contours in the $H_{0}-\eta$ plane for the $R_{\rm h}=ct$
Universe are shown in figure~6. The best-fit values for the simulated sample are
$H_{0}=65_{-0.63}^{+0.57}$ $(1\sigma)$ km $\rm s^{-1}$ $\rm Mpc^{-1}$ and
$\eta=0.048_{-0.011}^{+0.011}(1\sigma)$. These are similar to those in the
standard model, but not exactly the same, reaffirming the importance of reducing
the data separately for each model being tested. With $N=135$, our analysis of
the simulated sample shows that in this case the BIC would favor $\Lambda$CDM
over $R_{\rm h}=ct$ by an overwhelming likelihood of $99.7\%$ versus only $0.3\%$
(i.e., the prescribed $3\sigma$ confidence limit).

\begin{figure}[hp]
\centerline{\includegraphics[angle=0,scale=1.0]{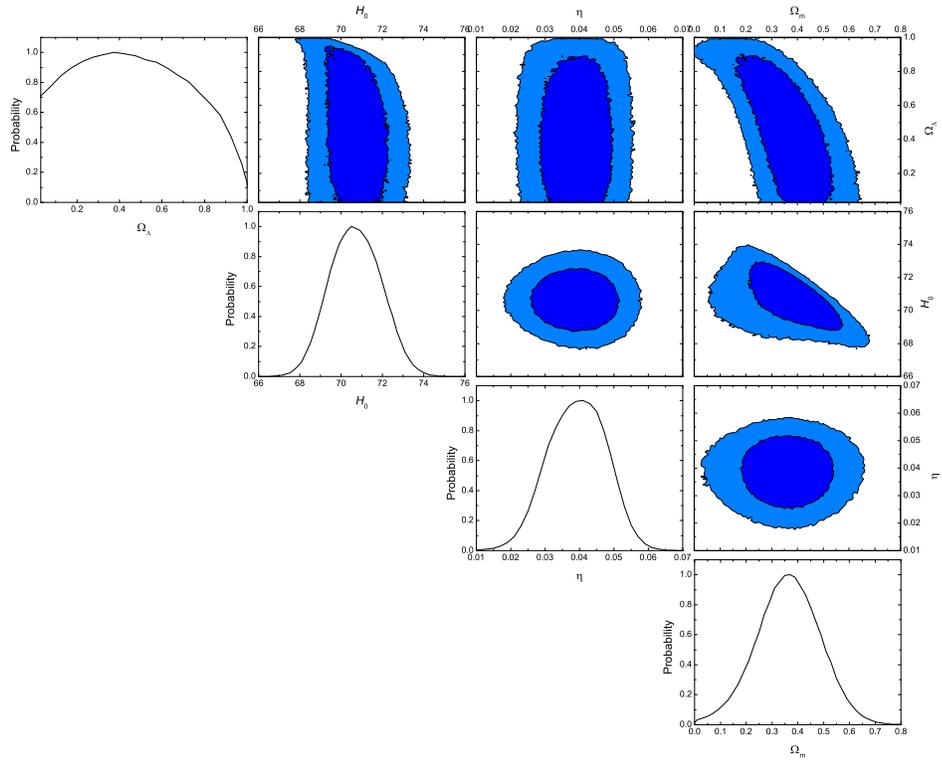}}
\vskip-0.2in
\caption{Same as Figure~3, except now with $\Lambda$CDM as the (assumed) background
cosmology. The simulated model parameters were $\Omega_{\rm m}=0.3$,
$\Omega_{\Lambda}=0.7$ and $H_{0}=70$ km $\rm s^{-1}$ $\rm Mpc^{-1}$.}
\end{figure}

\section{Conclusions}
The general agreement between theory and observation displayed in Figures~1
and 2 is promising, particularly since this work was based on the use of a
single isothermal sphere profile for the lens mass distribution. It would
be helpful to have additional information from which one may extract the
lens structure from individual images. Up to now, these approaches have
been used to optimize parameters in $\Lambda$CDM itself, but not for an
actual one-on-one comparison between competing cosmological models.
This must be done because the results we have presented here already
indicate a strong likelihood of being able to discriminate between models such as
$\Lambda$CDM and $R_{\rm h}=ct$. Such comparisons have already been made
using, e.g., cosmic chronometers (Melia \& Maier 2013), gamma-ray bursts
(Wei et al. 2013), and Type Ia SNe (Wei et al. 2014). The use of time-delay
lenses introduces yet another standard ruler, with systematics different
from those encountered elsewhere, thus providing an invaluable tool with
which to cross-check the outcomes of these other important tests.

\begin{figure}[hp]
\centerline{\hskip 0.5in\includegraphics[angle=0,scale=0.8]{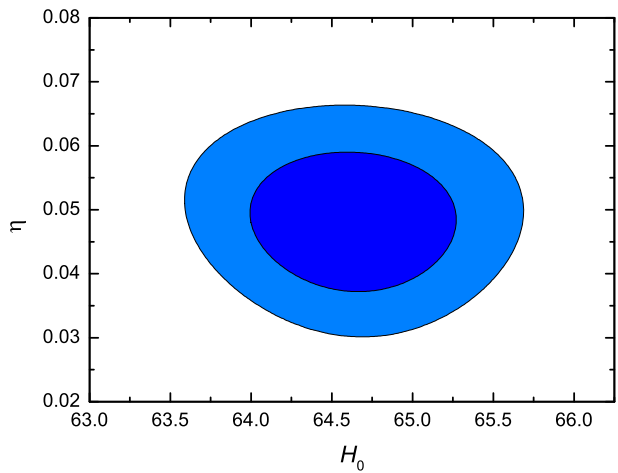}}
\vskip-0.2in
\caption{Same as Figure~4, except now with $\Lambda$CDM as the (assumed)
background cosmology.}
\end{figure}

But though time-delay lens observations currently tend to favor $R_{\rm h}=ct$
over $\Lambda$CDM, the known sample of such systems is still too small
for us to completely rule out either model. We have therefore considered
two synthetic samples with characteristics similar to those of the 12 known
systems, one based on a $\Lambda$CDM background cosmology, the other on
$R_{\rm h}=ct$. From the analysis of these simulated lenses, we have
estimated that a sample of about 150 lenses would be needed to rule out
$R_{\rm h}=ct$ at a $\sim 99.7\%$ confidence level if the real cosmology
is in fact $\Lambda$CDM, while a sample of at least 1,000 systems would
be needed to similarly rule out $\Lambda$CDM if the background cosmology
were instead $R_{\rm h}=ct$. The difference in required sample size
results from $\Lambda$CDM's greater flexibility in fitting the data, since
it has a larger number of free parameters. Such a level of accuracy may be
achievable with the successful implementation of surveys, such as DES, VST
ATLAS, and LSST.

\vskip-0.2in
\acknowledgments
We gratefully acknowledge helpful discussions with Gary Bernstein concerning the
Dark Energy Survey's expected image quality, and we thank the anonymous referee for
important suggestions to greatly improve this manuscript. This work is partially
supported by the National Basic Research Program (``973" Program)
of China (Grants 2014CB845800 and 2013CB834900), the National Natural Science Foundation
of China (grants Nos. 11322328 and 11373068), the One-Hundred-Talents Program,
the Youth Innovation Promotion Association, and the Strategic Priority Research Program
``The Emergence of Cosmological Structures" (Grant No. XDB09000000) of
the Chinese Academy of Sciences, and the Natural Science Foundation of Jiangsu Province.
F.M. is grateful to Amherst College for its support through a John Woodruff Simpson
Lectureship, and to Purple Mountain Observatory in Nanjing, China, for its hospitality
while part of this work was being carried out. This work was partially supported by grant
2012T1J0011 from The Chinese Academy of Sciences Visiting Professorships for Senior
International Scientists, and grant GDJ20120491013 from the Chinese State Administration
of Foreign Experts Affairs.

\end{document}